\documentclass[pdflatex,sn-mathphys-num]{sn-jnl}

\usepackage{graphicx}%
\usepackage{multirow}%
\usepackage{amsmath,amssymb,amsfonts}%
\usepackage{amsthm}%
\usepackage{mathrsfs}%
\usepackage[title]{appendix}%
\usepackage{xcolor}%
\usepackage{textcomp}%
\usepackage{manyfoot}%
\usepackage{booktabs}%
\usepackage{algorithm}%
\usepackage{algorithmicx}%
\usepackage{algpseudocode}%
\usepackage{listings}%
 \usepackage{subcaption}

\theoremstyle{thmstyleone}%
\theoremstyle{thmstyletwo}%

\theoremstyle{thmstylethree}%

\begin{document}

\title[IMUFace]{IMUFace: Towards Always-On 3D Facial Reconstruction via Earphone Inertial Sensing}

\author[1]{\fnm{Xianrong} \sur{Yao}}\email{ftxryao@mail.scut.edu.cn}
\author[1]{\fnm{Lingde} \sur{Hu}}\email{202320163218@mail.scut.edu.cn}
\author[1]{\fnm{Dong} \sur{She}}\email{ftdshe@mail.scut.edu.cn}
\author[2]{\fnm{Yincheng} \sur{Jin}}\email{yjin5@binghamton.edu}
\author[1]{\fnm{Yang} \sur{Gao}}\email{gaoyang2025@scut.edu.cn}
\author*[1]{\fnm{Zhanpeng} \sur{Jin}}\email{zjin@scut.edu.cn}

\affil*[1]{\orgdiv{School of Future Technology}, \orgname{South China University of Technology}, \country{China}}

\affil[2]{\orgdiv{Department of Computer Science}, \orgname{Binghamton University}, \country{United States}}

\abstract{Facial expression reconstruction technology offers considerable potential in areas like human-computer interaction, affective computing, and virtual reality. To tackle the privacy challenges and environmental constraints inherent in traditional camera-based systems, researchers have recently introduced ear-worn devices as a viable solution. Nevertheless, these methods still demand enhancements, particularly in aspects such as design appeal and energy efficiency, to achieve broader applicability and practical use. This paper introduces a system called IMUFace. It uses inertial measurement units (IMUs) embedded in wireless earphones to detect subtle ear movements caused by facial muscle activities, allowing for covert and low-power facial reconstruction. A user study involving 12 participants was conducted, and a deep learning model named IMUTwinTrans was proposed. The results show that IMUFace can accurately predict users' facial landmarks with a precision of 2.21 mm, using only five minutes of training data. The predicted landmarks can be utilized to reconstruct a three-dimensional facial model. IMUFace operates at a sampling rate of 30 Hz with a relatively low power consumption of 58 mW. The findings presented in this study demonstrate the real-world applicability of IMUFace and highlight potential directions for further research to facilitate its practical adoption.
}


\maketitle

\section{Introduction}
\label{sec:introduction}

Facial expressions play a crucial role in interpersonal communication, as they convey a wide range of information, including emotional states~\cite{bassili1979emotion, krumhuber2023role} and physical health conditions~\cite{argaud2018facial, roter2006expression}. As a result, facial expression reconstruction has become an essential tool for applications requiring facial analysis. This technology is fundamental to domains such as human-computer interaction (HCI), virtual reality (VR), affective computing, and healthcare. For instance, in VR, it improves users' real-time perception of emotions and facial expressions, creating a more immersive experience~\cite{kumari2015facial}. In metaverse platforms, it enables faster understanding and connection within virtual environments~\cite{cha2020real, frith2009role}. It also supports silent speech interfaces, allowing seamless communication between humans and computers~\cite{denby2010silent}.

Traditional methods for reconstructing facial expressions often relied on manual annotation or geometric modeling. These early approaches were labor-intensive, difficult to automate, and prone to practical limitations. However, with advancements in machine learning and computer vision, automated methods have significantly improved the field. Vision-based techniques~\cite{cootes2001active, cootes1995active, wang2020deep} enable precise facial landmark detection and high-quality animation generation. Nevertheless, their reliance on direct camera views introduces several challenges. These methods require users to face the camera under controlled conditions, making them less effective in environments with varying lighting, resolution, or occlusion. Additionally, the direct capture of facial imagery raises privacy concerns, particularly in dynamic or sensitive scenarios involving 3D head movements.

To address these drawbacks, researchers have shifted toward non-visual approaches for facial expression reconstruction. Wearable devices combined with multimodal sensors have emerged as a promising solution, offering greater flexibility and enhanced privacy. Examples include necklaces~\cite{chen2021neckface}, hats~\cite{bello2023inmyface}, glasses~\cite{xie2021acoustic, li2024eyeecho, li2024auglasses}, and ear-mounted devices~\cite{choi2022ppgface, song2022facelistener, wu2021bioface, li2022eario, zhang2023earphone}. Earlier designs, such as C-Face~\cite{chen2020c}, utilized miniature cameras mounted on the ear to capture expressions, providing a flexible deployment option. However, privacy concerns persisted due to the camera's proximity and mobility. Subsequent innovations like BioFace-3D~\cite{wu2021bioface} and EarIO~\cite{li2022eario} introduced non-visual 3D facial reconstruction with ear-mounted devices, though their designs sometimes caused discomfort due to protrusions or facial contact. More recent developments, such as EARFace~\cite{zhang2023earphone}, incorporated acoustic sensors into earplug designs, achieving higher usability ratings. Nonetheless, the high power consumption of acoustic sensors remains a limitation for extended use. Therefore, creating a portable, comfortable, and energy-efficient system capable of real-time facial expression reconstruction continues to be a significant challenge.

To address these challenges, we present IMUFace, an earphone platform equipped with two IMUs for precise facial reconstruction. These IMUs are highly sensitive, allowing them to capture subtle ear movements caused by facial muscle activity quickly and accurately.

\begin{figure}
	\centering
	\includegraphics[width=1\textwidth]{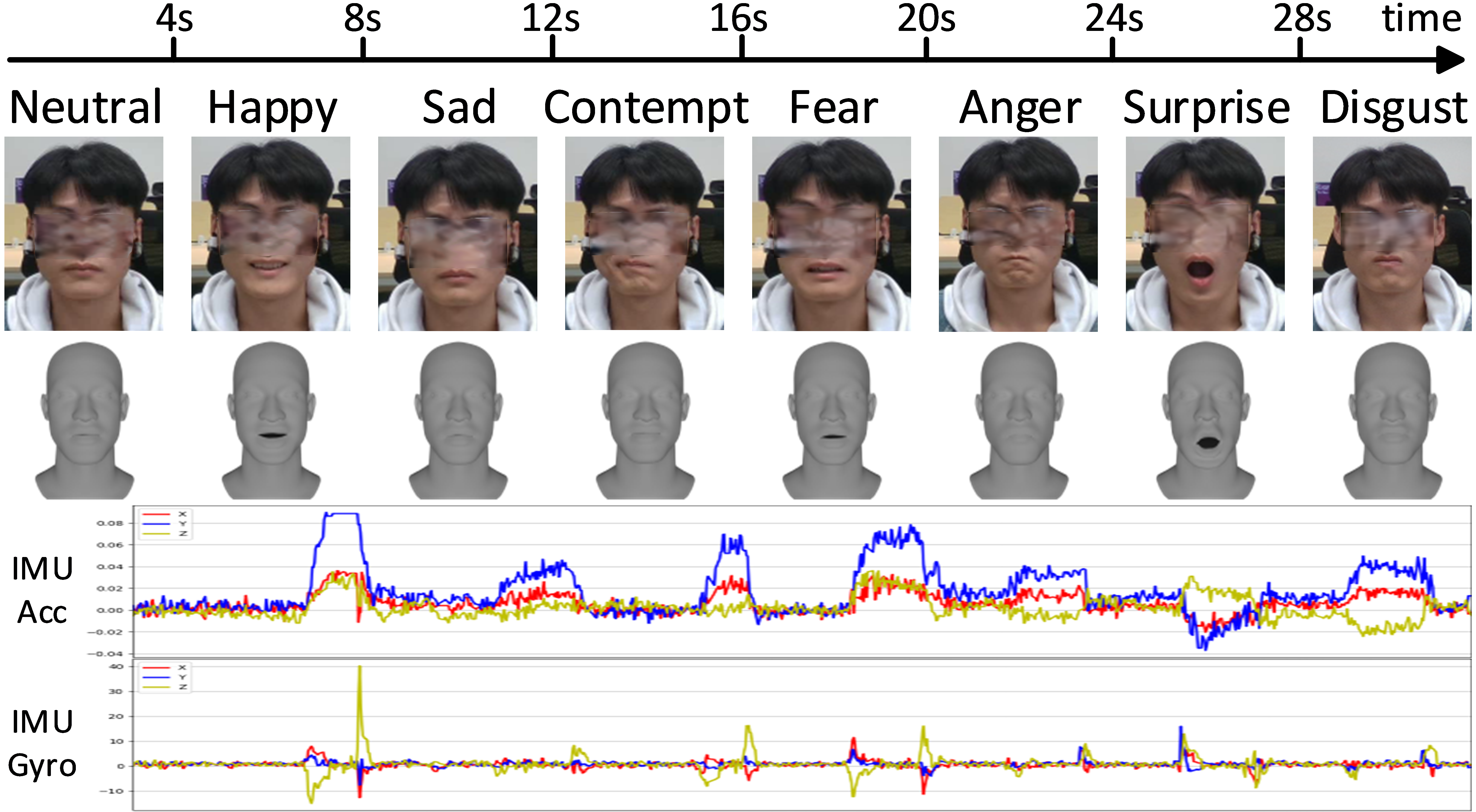}
	\caption{Demonstration of continuous 3D facial reconstruction of IMUFace.}
	\label{fig:1}
\end{figure}

We developed a deep learning model called IMUTwinTrans to achieve 3D facial reconstruction beyond the limitations of cameras. This model, based on the ConvTransformer architecture, is capable of simultaneously leveraging the temporal and frequency features of IMU signals to establish the correspondence between IMU signals and facial landmarks. Additionally, the model adopts a lightweight design to meet the requirements for real-time performance. During testing, the well-trained model can localize facial landmarks directly from IMU signals without visual input. The predicted facial landmarks are then filtered and fitted into a generic 3D head model to generate continuous 3D facial animation. Fig.~\ref{fig:1} showcases 3D facial reconstruction by the IMUFace system and, together with the real-time IMU gyroscope signals (X, Y, Z axes), highlights how subtle angular changes from facial muscle movements produce distinct peaks and troughs that correspond to different expressions, confirming both the sensor’s high sensitivity and the physical meaning of the IMU data for deep learning–based facial reconstruction.

The main contributions of this study are as follows:
\begin{enumerate}
    \item We presented the first attempt to use IMU sensor signals on wireless earphones for facial landmark tracking to achieve user-friendly, continuous 3D facial expression reconstruction.
    \item We designed a lightweight deep learning model, IMUTwinTrans, based on the ConvTransformer architecture, to achieve 3D facial reconstruction by leveraging both temporal and frequency features of IMU signal while meeting real-time performance requirements.
    \item We optimized the power consumption of the system, enabling it to operate at a sampling rate of 30 Hz with a power consumption of only 58 mW.
\end{enumerate}

\section{Related Work}

The field of non-wearable 3D face reconstruction has advanced significantly, driven by developments in morphable models, photometric stereo methods, and deep learning approaches. These methods address challenges such as variations in pose, illumination, and image quality, enabling high-fidelity reconstruction from unconstrained inputs.

Morphable Models serve as a foundational tool for compactly representing facial geometry and texture. Roth et al.~\cite{roth2016adaptive} introduced an adaptive 3DMM-based method for reconstructing faces from unconstrained photo collections. Their approach utilized a coarse-to-fine fitting strategy, enhancing robustness against low-quality images and diverse imaging conditions. An earlier work~\cite{roth2015unconstrained} proposed an algorithm for generating accurate 3D models from 2D images without relying on metadata, incorporating Laplace editing to improve resilience to pose and illumination variations.

Photometric Stereo techniques have been pivotal in capturing surface normals from multiple viewpoints, contributing to detailed facial reconstructions. For instance, Roth et al.~\cite{roth2015unconstrained} integrated photometric stereo with landmark constraints to achieve higher reconstruction accuracy. Xie et al.~\cite{xie2021scifi} demonstrated a practical smartphone-based method using calibrated screen lighting, addressing challenges like hand jitter through alignment and outlier removal techniques.

Deep Learning has transformed 3D face reconstruction, enabling high-fidelity results from single images. Gecer et al.~\cite{gecer2019ganfit} combined Generative Adversarial Networks (GANs) with Deep Convolutional Neural Networks (DCNNs) in GANFIT, achieving photorealistic reconstructions of facial shape and texture. Fast-GANFIT~\cite{gecer2021fast}, their subsequent work, introduced a self-supervised regression-based initialization scheme, improving speed and stability.

\subsection{3D Face Reconstruction with Different Models}

The 3D Morphable Model (3DMM)~\cite{blanz2023morphable} remains a cornerstone in 3D face reconstruction. While traditional methods optimize coefficients for reconstruction, which is computationally expensive, recent advancements utilize deep learning to directly predict these coefficients, offering greater efficiency and precision.

Convolutional Neural Networks (CNNs) are widely employed in this field. Richardson et al.~\cite{richardson2017learning} developed an end-to-end CNN framework for reconstructing facial geometry in a coarse-to-fine manner. Tian et al.~\cite{tian2018landmark} presented a CNN-based method for reconstructing 3D faces from two images, demonstrating superior performance on benchmark datasets.

Encoder-Decoder Architectures have also gained prominence. Ji et al.~\cite{ji2020view} proposed a multi-task encoder-decoder framework based on a Siamese network to ensure view-consistent reconstruction of facial shapes and textures from single images. Similarly, Li et al.~\cite{li2020novel} enhanced accuracy by employing a dual-pathway encoder-decoder network that separates facial attributes.

Transformer Models have recently achieved remarkable progress. Chen et al.~\cite{chen2022transformer} utilized a conditional Generative Adversarial Network (cGAN) combined with a mesh transformer for cross-domain face synthesis, introducing a reprojection consistency loss for self-supervised learning. Yaermaimaiti et al.~\cite{yaermaimaiti2024research} integrated ResNet and Transformer modules, leveraging self-attention mechanisms to improve parameter prediction and reconstruction quality.

\subsection{Non-wearable 3D Face Reconstruction}

Traditional methods in 3D face reconstruction rely on external imaging devices like RGB cameras to process extensive photo collections. Roth et al.~\cite{roth2016adaptive} proposed a method for reconstructing personalized 3D face models from photos captured under varying poses and lighting conditions. Their approach employed a 3D Morphable Model and a novel photometric stereo formulation to achieve high-quality results. Furthermore, Roth et al.~\cite{roth2015unconstrained} introduced an algorithm that integrates photometric stereo with Laplace editing to handle unconstrained 2D image collections effectively.

\subsection{Wearable 3D Face Reconstruction}

Recent research has focused on wearable devices for 3D face reconstruction, enabling real-time applications with minimal interference. Chen et al.~\cite{chen2020c} designed an ear-mounted device equipped with two miniature cameras to capture subtle facial contour changes. Their deep learning model continuously outputs facial feature points, enabling real-time expression reconstruction. Aoki et al.~\cite{aoki2021facerecglasses} developed a glasses-shaped device featuring two cameras and three mirrors to capture facial expressions unobtrusively, achieving high accuracy in emotion recognition.

Multimodal Approaches have further enhanced wearable devices. Wu et al.~\cite{wu2021bioface} introduced a single-earpiece biosensing system that tracks facial landmarks and generates 3D animations from biosignals, achieving performance comparable to camera-based methods. Li et al.~\cite{li2024auglasses} utilized IMUs in smart glasses to estimate action unit intensities, enabling real-time, low-power facial reconstruction. Zhang et al.~\cite{zhang2023earphone} employed in-ear acoustic sensors to track facial landmarks by detecting ear canal shape changes, providing a privacy-preserving solution for 3D facial reconstruction.

\section{Background}
This section provides an overview of the foundational concepts, covering: (1) the role of facial muscles in generating expressions and (2) the operational principle of IMUFace.

\begin{figure}
	\centering
	\includegraphics[width=.8\textwidth]{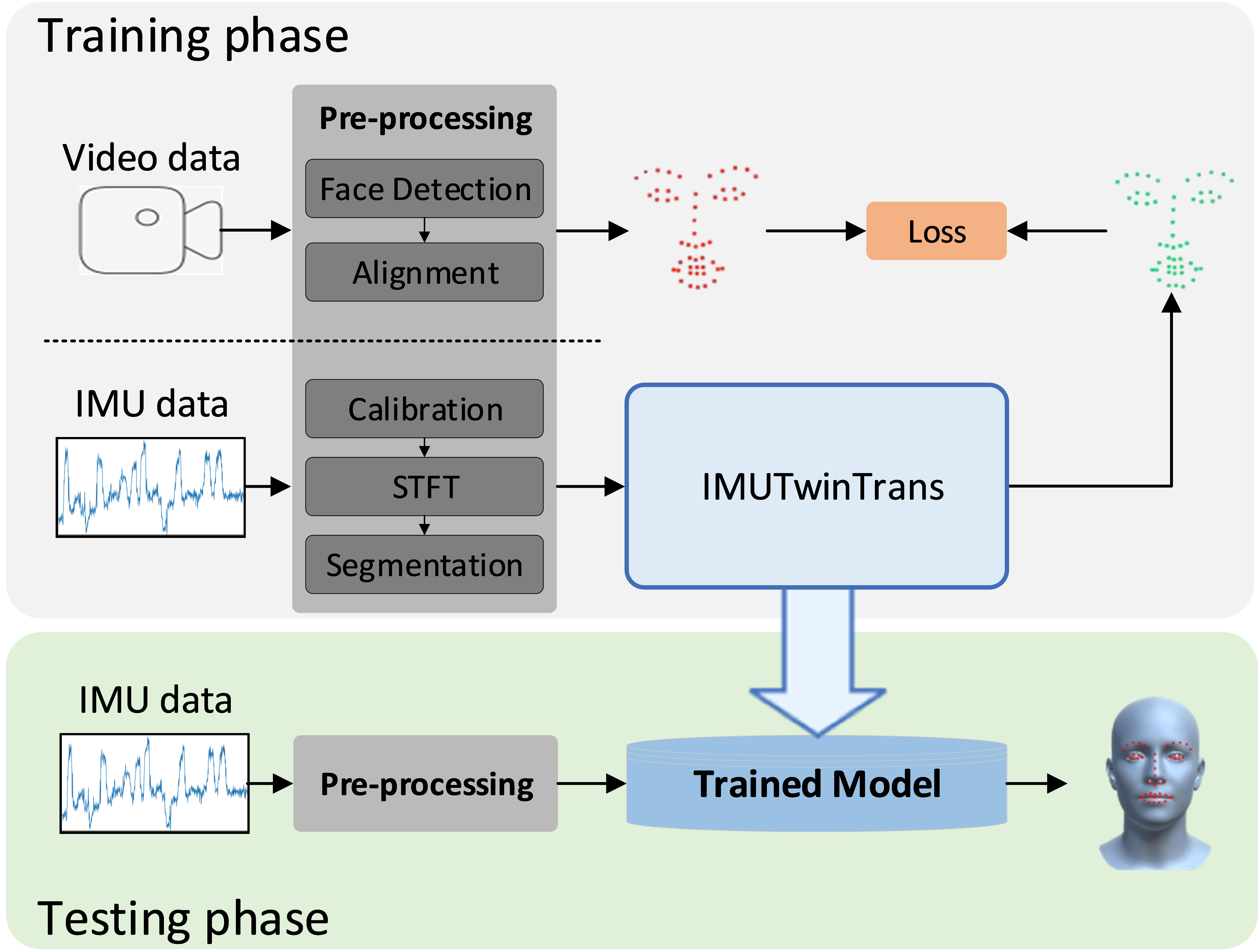}
	\caption{Overview of the IMUFace system.}
	\label{fig:4}
\end{figure}

\subsection{Facial Muscles and Expressions}

Facial expressions play an essential role in human communication by conveying emotions and intentions that can be observed and interpreted by others. The human face contains 43 distinct mimetic muscles, which collectively enable the production of over 10,000 unique expressions. These muscles are categorized into three primary groups: orbital, nasal, and oral. The orbital muscles are responsible for eyelid movements, the nasal muscles manage actions around the nose, and the oral muscles control the lips and mouth. The coordination of these groups enables the depiction of a broad range of emotions, such as happiness, sadness, surprise, and fear.

The intricate nature of facial expressions stems from the synchronized activity of these muscle groups, which exert forces on the skin to form visible emotional cues. For instance, the frontalis muscle elevates the eyebrows during expressions of surprise, while jaw movements in amazement or joy involve the temporalis and lateral pterygoid muscles.

\subsection{Sensing Principle}

The interplay between facial muscle activity and associated anatomical changes provides the basis for innovative sensing methods that indirectly capture facial expressions. Muscle contractions during expressions not only generate visible changes but also induce subtle structural alterations in nearby areas, such as the ear canal, due to the dynamics of interconnected tissues and bones.

For example, jaw movements that occur during expressions like joy or surprise influence the mandibular condyle, causing changes in the ear canal's shape. Similarly, the temporalis muscle, a key component of the head's musculature, transmits mechanical effects of facial expressions to the ear region.

By embedding an Inertial Measurement Unit (IMU) in an ear-mounted device, such as a wireless earbud, these changes can be detected. As facial expressions deform the ear canal, they alter the earbud's position and orientation. The IMU captures these variations using accelerometers and gyroscopes, converting them into measurable signals. Each facial expression produces a distinct motion pattern, allowing the IMU to differentiate expressions based on the recorded data. Advanced machine learning algorithms process these signals to identify specific expressions, enabling unobtrusive and practical facial expression recognition. This approach capitalizes on the natural linkage between facial musculature and ear canal dynamics, offering a novel solution for wearable expression recognition systems.

\section{Method}
In Fig.~\ref{fig:4}, we present an overview of the architecture of our proposed IMUFace system, which is divided into two distinct phases: a training phase and a testing phase.

The training phase aims to teach a deep learning model to map IMU signals to facial movements using two synchronized data streams: video and IMU. The video stream provides ground truth by detecting and aligning faces in each frame, from which 2D facial landmarks are extracted. Meanwhile, the IMU stream, collected from earphone-embedded sensors, captures subtle facial muscle movements. This data undergoes a three-step preprocessing pipeline: a quick calibration to adapt to individual facial structures, a Short-Time Fourier Transform (STFT) to convert signals into the frequency domain, and windowed segmentation for temporal analysis. The processed IMU data is then fed into our transformer-based model, IMUTwinTrans, which predicts facial landmarks. A loss function measures the difference between the predicted and ground-truth landmarks, enabling the model to optimize its parameters through backpropagation.

Once training is complete, the system enters the testing phase for inference. In this phase, only the IMU data stream is required. The incoming IMU data undergoes the same preprocessing steps as in the training phase (calibration, STFT, and segmentation). This processed data is then fed into the fully trained model. The model, having learned the correlation between IMU signals and facial expressions, directly outputs the estimated facial landmarks in real-time. As the final step, these predicted landmarks are fitted to the parameters of the FLAME model\cite{li2017learning} to reconstruct and animate a realistic 3D facial avatar.

\begin{figure}
	\centering
	\includegraphics[width=.55\textwidth]{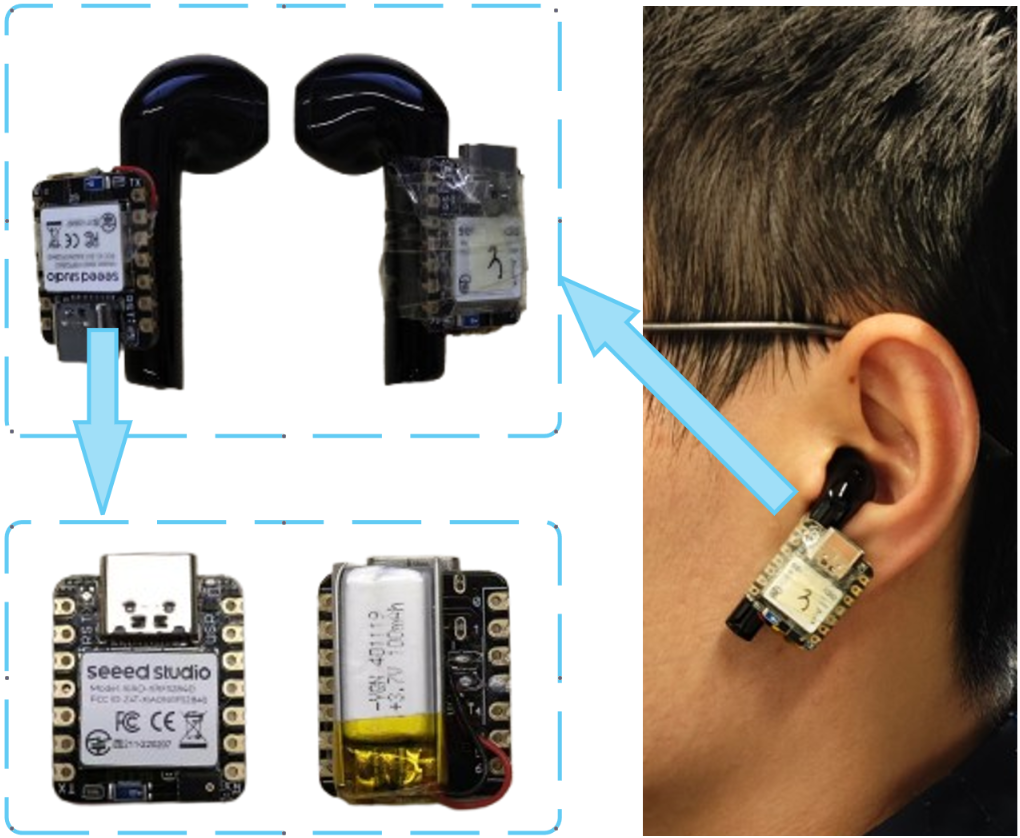}
	\caption{Device Prototype: Composed of a pair of earbuds and XIAO nRF52840 Sense MCU.}
	\label{fig:2}
\end{figure}

\subsection{Hardware Prototype Design}
We constructed an earphone-based, proof-of-concept prototype capable of capturing IMU signals generated by facial expression movements. It primarily consists of two parts: a tiny sensor board and an earphone frame. We selected the XIAO nRF52840 Sense as the sensor board, with a Bluetooth chip and a 6-axis IMU chip. The Bluetooth chip supports Bluetooth 5.0 for data transmission and directly interfaces with the IMU chip, extracting data using the I2C protocol. The sensor board measures 21.0 mm × 17.5 mm. It provides a compact and efficient solution with low power consumption, meeting our requirements perfectly. 

The earphone frame we chose is a standard wireless Bluetooth earphone found in daily life, which can be any standard earphone, ensuring the broad applicability of the system. We stream the 6-axis IMU data from both earphones at a sampling rate of 30 Hz with a 16-bit resolution, ensuring the comprehensiveness of the collected muscle information. The schematic diagram of the device is shown in Fig.~\ref{fig:2}. First, we attach the battery to the back of the XIAO nRF52840 Sense, then mount it on the earphone. Actually, many modern wireless earphones/earbuds are already equipped with the IMU module inside, which could be easily adapted for deploying the IMUFace approach. 

\subsection{Data Processing Pipeline}

\subsubsection{Calibration} 
Since different users may have different neutral or natural head orientations, we ask users to wear the earbuds and keep their heads naturally upright before starting the video recording. We then record 4 seconds of IMU data from all channels and take the sample mean of these values, referred to as the offset. To compensate for the drift in gyroscope measurements, we adjust the gyroscope readings of each frame by the calculated offset.

\subsubsection{Signal Synchronization and Filtering}
To ensure synchronization between the two modalities of data streams, we record timestamps for both video data and IMU sensor data. Using the starting timestamp of the video data as a reference, we find the closest corresponding timestamp in the IMU sensor data. Since the sampling rate of the IMU sensor data fluctuates around 30 Hz, we resample the IMU sensor data to 30 Hz. Although our hardware device includes a built-in bandpass filter for reading IMU sensor data, we apply a high-pass filter with a cutoff frequency of 0.1 Hz to improve the signal-to-noise ratio.

\begin{figure}[!t]
\centering
\captionsetup[subfigure]{justification=centering}
\subfloat[]{
		\includegraphics[width=0.35\linewidth]{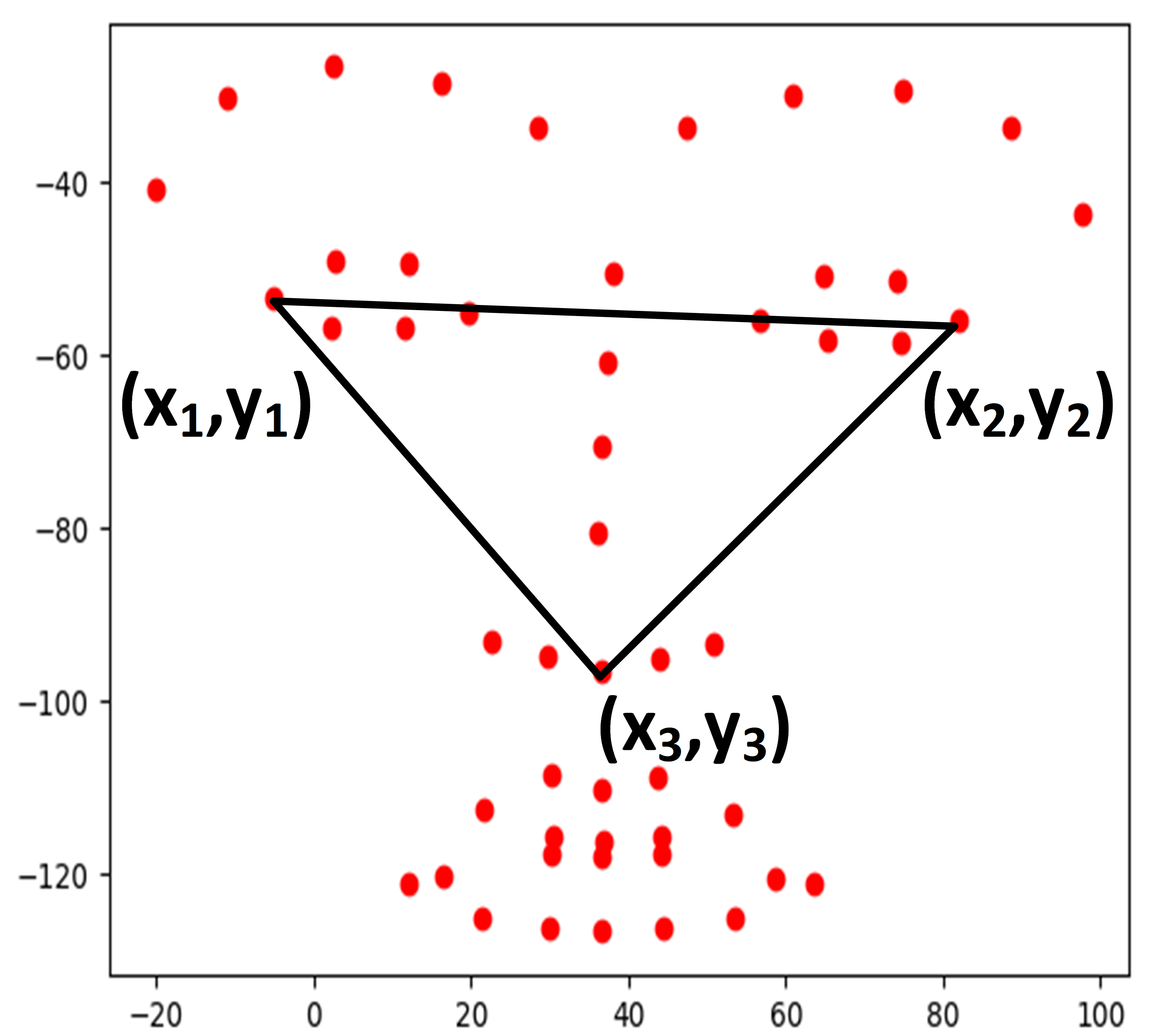}}
\subfloat[]{
		\includegraphics[width=0.35\linewidth]{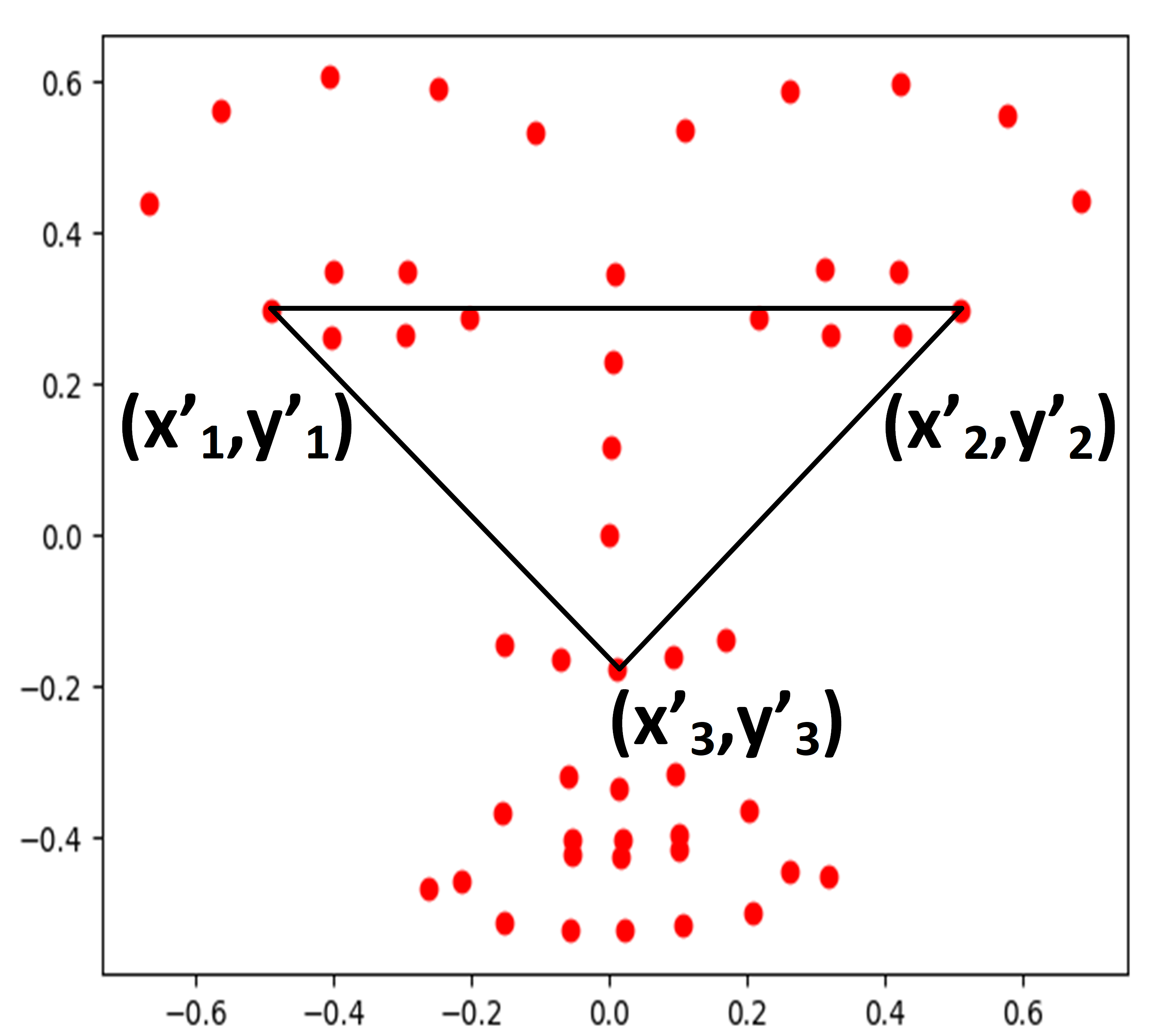}}
\caption{Illustration of facial landmark alignment. (a) Before landmark alignment. (b) After landmark alignment.}
\label{fig:3}
\end{figure}

\subsubsection{Signal Normalization}
\label{sec:SignalNormalization}
We utilize the open-source tool OpenFace \cite{8373812} to extract facial landmarks from facial videos. However, the positions of these landmarks can be affected by factors such as head position, pose variations, and changes in camera angle and placement. Following the approach used in the 300W dataset \cite{6755925}, we standardize the facial landmarks to a consistent reference frame through a series of transformations.

First, we select the nose tip as the origin and shift all landmarks accordingly, as defined in Eq.~\eqref{eq_1}:
\begin{equation}\label{eq_1}
    P_{i}' = (x_i - x_{\text{nose}}, y_i - y_{\text{nose}})
\end{equation}
where \( P_i = (x_i, y_i) \) denotes the original landmark coordinates, and \( (x_{\text{nose}}, y_{\text{nose}}) \) represents the position of the nose tip.

Next, we perform a coordinate rotation to align the \( x \)-axis with the direction from the left outer eye corner to the right outer eye corner. The rotation angle \( \theta \) is computed using Eq.~\eqref{eq_2}:
\begin{equation}\label{eq_2}
    \theta = \arctan\left(\frac{y_{\text{right}} - y_{\text{left}}}{x_{\text{right}} - x_{\text{left}}}\right)
\end{equation}
where \( (x_{\text{left}}, y_{\text{left}}) \) and \( (x_{\text{right}}, y_{\text{right}}) \) correspond to the coordinates of the outer corners of the left and right eyes, respectively.

The rotated landmark coordinates \( P_i'' \) are then obtained by applying the transformation in Eq.~\eqref{eq_3}:
\begin{equation}\label{eq_3}
P_{i}'' = 
\begin{bmatrix}
\cos\theta & -\sin\theta \\
\sin\theta & \cos\theta
\end{bmatrix}
\begin{bmatrix}
x_i' \\
y_i'
\end{bmatrix}
\end{equation}

To ensure uniform scaling, we normalize the distance between the two outer eye corners to 1. The normalization factor \( d \) is calculated according to Eq.~\eqref{eq_4}:
\begin{equation}\label{eq_4}
d = \sqrt{(x_{\text{right}}'' - x_{\text{left}}'')^2 + (y_{\text{right}}'' - y_{\text{left}}'')^2}
\end{equation}

Finally, the normalized landmark coordinates \( P_i''' \) are obtained by applying Eq.~\eqref{eq_5}:
\begin{equation}\label{eq_5}
P_{i}''' = \frac{P_{i}''}{d}
\end{equation}

Fig.~\ref{fig:3} shows the facial landmark coordinates before and after alignment.

\subsubsection{Short-Time Fourier Transform} 
Each facial expression movement is associated with unique frequency components. Since the STFT better isolates noise components and enhances the clarity of extracted features from time-series data~\cite{zhang2022force}, we use STFT to convert time-domain signals into time-frequency domain spectrograms.  The STFT of a signal \( x(t) \) is computed as Eq.~\eqref{eq_6}:

\begin{equation}\label{eq_6}
X(t, f) = \int_{-\infty}^{\infty} x(\tau) w(\tau - t) e^{-j2\pi f \tau} \, d\tau
\end{equation}
where \( w(\tau - t) \) is a window function centered at \( t \), \( f \) is the frequency, and \( X(t, f) \) represents the time-frequency representation of the signal. Frequency-domain features improve the training of our model. We apply STFT to each channel of the IMU data with a window length of 30, a step size of 1, and 32 frequency components. Considering the model complexity, we concatenate six spectrograms (sharing the same time axis) to form a single feature representation input. The concatenation is along the frequency axis, resulting in a final frequency-domain feature dimension of \( 12 \times 17 = 204 \), with a length of IMU signal.

\begin{figure}
    \centering
    \includegraphics[width=0.8\linewidth]{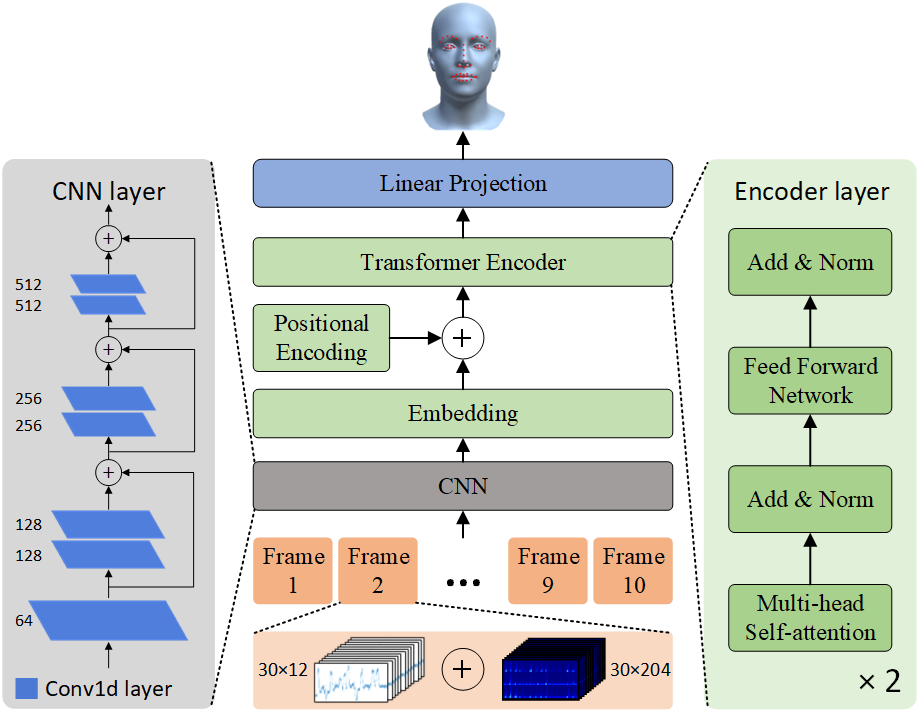}
    \caption{Illustration of the model structure.}
    \label{fig:42}
\end{figure}

\subsubsection{Data Segmentation}
We collect IMU signal data with 12 channels and extract frequency-domain features for each channel. Both the time-domain data and the extracted frequency-domain features are used as inputs to the model. To help the model capture temporal relationships when predicting facial landmarks, we use ten consecutive frames of processed IMU data as one input segment.

\subsection{Model Architecture and Training Strategy}
In this study, we designed a deep-learning model called IMUTwinTrans. This model, based on a ConvTransformer architecture, leverages both temporal and frequency features of IMU signals to establish the correspondence between IMU data and facial landmarks, as illustrated in Fig.~\ref{fig:42}. The model input consists of 10 consecutive frames of IMU data that include both time-domain and frequency-domain features. Specifically, the time-domain signal for each frame is arranged as a $30\times12$ matrix. The frequency-domain features are obtained via a Short-Time Fourier Transform (STFT), resulting in a $30\times204$ matrix. By concatenating the time-domain and frequency-domain features, each frame forms a combined feature matrix of $30\times216$. The input data are first processed by a Convolutional Neural Network (CNN) for feature extraction, followed by a Transformer encoder to capture temporal dependencies. Finally, a linear projection layer outputs the predicted coordinates of 51 facial landmarks, represented as a $1\times102$ vector.

\subsubsection{CNN-based Feature Extraction}
We utilize a CNN to extract robust features from the input data frames. The architecture consists of sequentially stacked 1D convolutional layers with residual connections to enhance feature representation. The network begins with an initial convolutional block followed by three residual blocks that progressively increase the number of channels from 64 to 512 while maintaining a consistent kernel size of 3 and a stride of 1. This configuration allows the network to capture both fine-grained and hierarchical information. To mitigate overfitting, we incorporate dropout layers with a probability of 0.15 after each convolutional layer.

Additionally, each residual block includes shortcut connections to preserve gradient flow during training, improving model stability and convergence. Batch normalization is applied after each convolution to ensure faster training and better generalization. The input frames of size 10×216 are passed through this network, yielding an output of size 10×512, effectively capturing informative features for downstream tasks.

\subsubsection{Transformer Structure}
We perform value embedding and position encoding on the feature data derived from the CNN output (size $10 \times 512$) and feed it into a two-layer Transformer Encoder. The encoder includes a self-attention module within the multi-head attention layers, a multilayer perceptron, and normalization layers. Residual connections are incorporated at both the multi-head attention and multilayer perceptron layers to boost the convergence of the machine learning model. Regarding model parameters, we set $d_{model}$ to 512, $n_{head}$ to 4, $d_{ff}$ to 1024, and the dropout ratio to 0.1 in the encoder.

\subsubsection{Loss Function}
In the task of facial landmark prediction, standard MSE and MAE loss functions treat all errors equally, regardless of their magnitude. In particular, MSE heavily penalizes large errors while neglecting small ones, resulting in important but less active landmarks (such as pupils compared to lips) not receiving sufficient attention during training. To address this issue, we adopt the Wing Loss function~\cite{feng2018wing}, which applies a relatively smooth treatment to small errors while assigning higher weights to large errors. This helps the model predict points closer to the true values more accurately without overly penalizing the model for a few large errors. The loss for each facial landmark is defined as (Eq.~\eqref{eq_7}): 

\begin{equation}\label{eq_7}
\text{Wing Loss}(x_i)=\begin{cases}w\log(1+\frac{|x_i|}{\epsilon}),&\text{if}|x_i|<w\\|x_i|-C,&\text{if}|x_i|\ge w\end{cases}
\end{equation}
where $|x_i|$ is the L2 distance between the ground truth and the reconstructed coordinate for the $i$ landmark. $w$ represents the threshold of the small error, which is set to 20 in our case. $\epsilon$ means the curvature in the small error range, and $C=w-wln(1+w/\epsilon)$ which links the linear part and non-linear part together.

\subsubsection{Optimization}
In addition, the network is trained using the Adam optimizer. The learning rate scheduler uses a Warmup + Cosine Annealing strategy. Initially, during the Warmup phase, the learning rate gradually increases to stabilize gradient updates and prevent divergence. This is followed by the Cosine Annealing phase, where the learning rate smoothly decreases following a cosine curve, allowing for more precise parameter adjustments and reducing overfitting. This strategy accelerates convergence while improving the model’s generalization.

\subsection{3D Reconstruction}
To enhance system usability and simplify the modeling process, we aim to use a compact head model that balances reduced complexity with sufficient detail to produce expressive facial animations.

The IMUFace system adopts the FLAME model (2020 version) for 3D facial representation. Specifically, the 3D reconstruction is primarily carried out using the FLAME PyTorch framework. FLAME is a statistical head model built from a dataset of approximately 33,000 3D scans. It efficiently captures complex facial geometries and expressions through a combination of identity PCA space, rotational degrees of freedom for the neck, jaw, and eyes, pose-dependent corrective blend shapes, and global expression blend shapes. The model is mathematically represented as (Eq.~\eqref{eq_8}):

\begin{equation}\label{eq_8}
FLAME(\vec{\beta},\vec{\theta},\vec{\psi}):\mathbb{R}^{|\vec{\beta}|\times|\vec{\theta}|\times|\vec{\psi}|}\to\mathbb{R}^{3N}
\end{equation}

A series of optimization steps is implemented to reconstruct a 3D face model that accurately reflects the user's expressions. The process begins with camera calibration, where the parameters for scale, rotation, and translation are optimized to minimize the L2 distance between the 2D projections of the 3D model's facial landmarks and the user's facial landmarks. Following this, the FLAME model parameters—covering shape ($\vec{\beta}$), pose ($\vec{\theta}$), and expression ($\vec{\psi}$)—are fine-tuned to further reduce the L2 distance between the projected landmarks of the 3D model and the user's facial landmarks. This L2-distance-based optimization method enables the model to fully adapt and reconstruct a unique shape that closely matches the target user's facial features.

Using the optimized FLAME parameters, the user's 3D face model can then be rendered with high accuracy. While we customize the shape parameters for each user to capture their unique head geometry, we uniformly apply FLAME's default texture parameters. This decision was primarily driven by our application's focus on the dynamic representation of facial expressions and the accuracy of geometric shapes. The default texture provides a sufficient basic visual effect while avoiding the additional complexity and data acquisition costs associated with obtaining and fitting individual textures. Furthermore, a unified texture simplifies the rendering process, contributing to the system's consistency and robustness.

\section{Performance Evaluation}

To validate the effectiveness of our system, we conducted extensive and detailed user studies to collect training data and evaluate the performance of our system in facial reconstruction. We also performed a series of micro-benchmark tests using the trained model to further ensure the system's reliability and better understand its capabilities. These tests were designed to examine the system's performance across various scenarios.

\subsection{Experimental Methodology}

\subsubsection{Data Collection}
In the study, we recruited 12 participants to evaluate IMUFace's performance. The participants included nine males and three females from the local institution, aged 18 to 28. The experimental equipment comprises the IMUFace prototype, a Hikvision camera, and a desktop computer. Participants were asked to wear the prototype of the IMUFace wireless earphones, which were connected to the computer via Bluetooth. \textcolor{red}{This study was reviewed and approved by the Ethics Review Committee (ERC) at the South China University of Technology.}


To evaluate the performance of tracking 51 facial landmarks, we focused on seven universal facial expressions of emotion involving happiness, sadness, anger, surprise, fear, disgust, and contempt. Before the experiment, we showed the participants examples of the seven facial expressions and had them repeatedly practice making the corresponding facial expressions while wearing our implemented IMUFace prototype. 

Each expression was separated by a neutral facial expression (e.g., a relaxed facial expression). To assist participants with their data collection, seven pictures of the corresponding faces were displayed on a screen for them to imitate. The pace and extent to which each expression was performed were not controlled throughout the experiments. While the users performed the facial motion, IMU data from both earphones were collected at a sampling rate of 30 Hz. To validate the accuracy of IMUFace, the built-in camera recorded videos of the user's face on the computer at 720p resolution and 30 fps.

Each participant completed 12 sessions of the experiment. In each session, the participants were required to make seven expressions, which were prompted in a random order. Each expression was made twice. Each data collection session lasted 60 seconds, and the participants took adequate rest between sessions. The earphone prototype was removed and remounted across sessions to validate the robustness to natural changes in sensor positions during daily usage. 

\subsubsection{Experimental Settings}
The training was performed on a desktop with an Intel i9-12700K CPU, 32 GB of RAM, and an Nvidia GTX 4090 GPU. We used the Adam optimizer with a learning rate of 0.001. To avoid overfitting issues that may happen in the training process, we added dropouts with a parameter of 0.1 following each ReLU activation.

\subsubsection{Evaluation Metric}
The IMUFace system is evaluated primarily by two metrics~\cite{wu2021bioface}: 

1) \textbf{Mean Absolute Error (MAE)}: the absolute difference between the predicted facial landmarks and the ground-truth facial landmarks. It is defined as: $MAE=\frac{1}{N}\times\sum\|g-r\|\times\frac{d_{real}}{d_{norm}}$, where $g$ and $r$ denote the facial landmarks extracted from the ground truth of vision and IMUFace in a normalized coordinate frame. ${d_{real}}$ denotes the actual outer corner distance in millimeters(we measured ${d_{real}}$ for each user who participated in the study). ${d_{norm}}$ denotes the outer corner distance in the normalized coordinate frame. From Section~\ref{sec:SignalNormalization}, we know that ${d_{norm}} = 1$. This provided the MAE in units of millimeters, which is the primary metric used to evaluate IMUFace. 

2) \textbf{Normalized Mean Error (NME)}: the average error between the predicted facial landmarks and the ground truth facial landmarks, represented by $NME=\frac{1}{N}\times\sum\frac{\|g-r\|}{d_{norm}}\times100\%$, where $g$ and $r$ denote the facial landmarks extracted from the ground truth of vision and IMUFace in a normalized coordinate frame. ${d_{norm}}$ denotes the outer corner distance in the normalized coordinate frame. From Section~\ref{sec:SignalNormalization}, we know that ${d_{norm}} = 1$. 

\begin{figure*}[!t]
\centering
\captionsetup[subfigure]{justification=centering}
\subfloat[]{
		\includegraphics[width=0.38\textwidth, height=0.2\textheight]{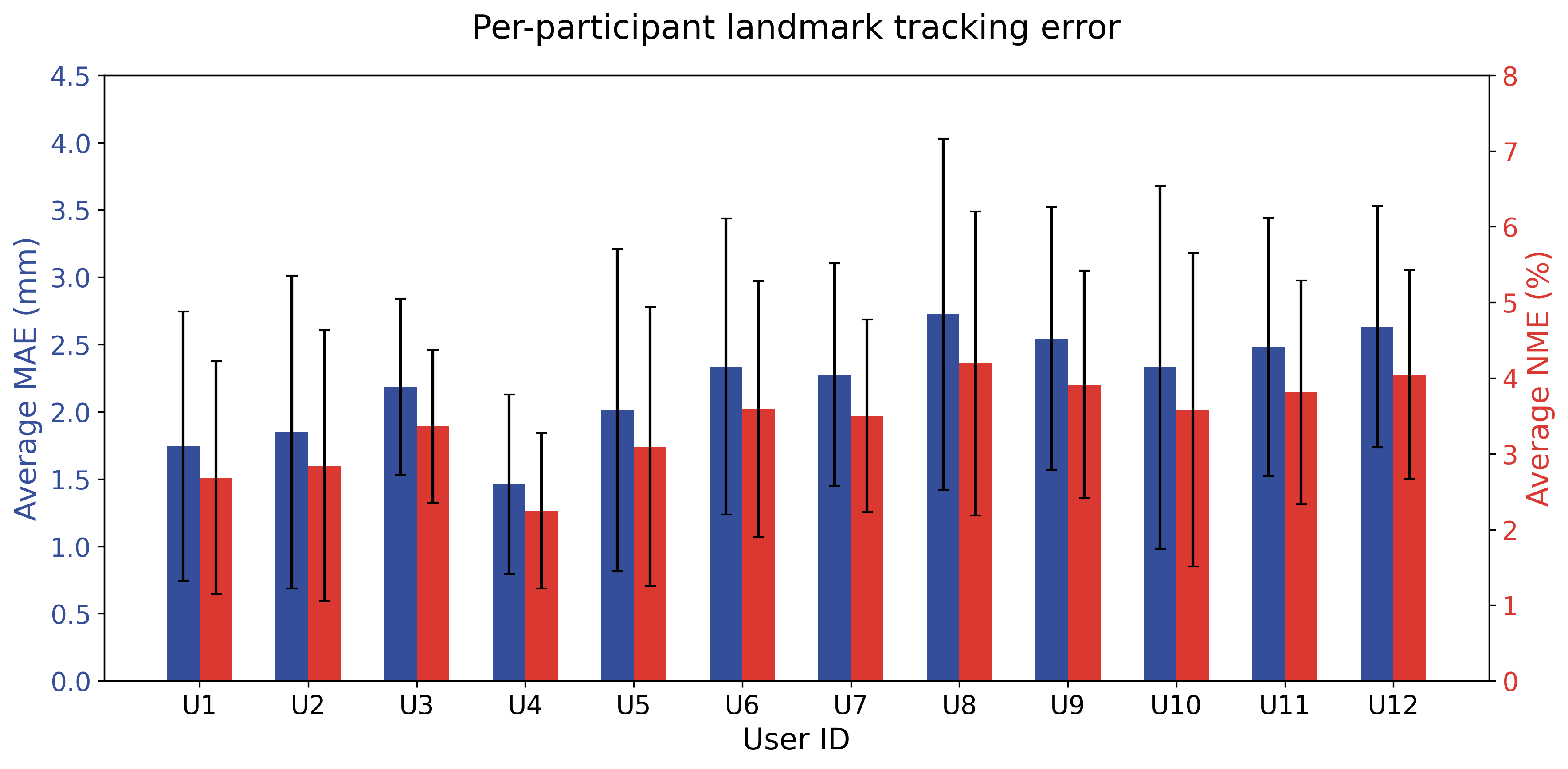}} \hfill
\subfloat[]{
		\includegraphics[width=0.27\textwidth, height=0.19\textheight]{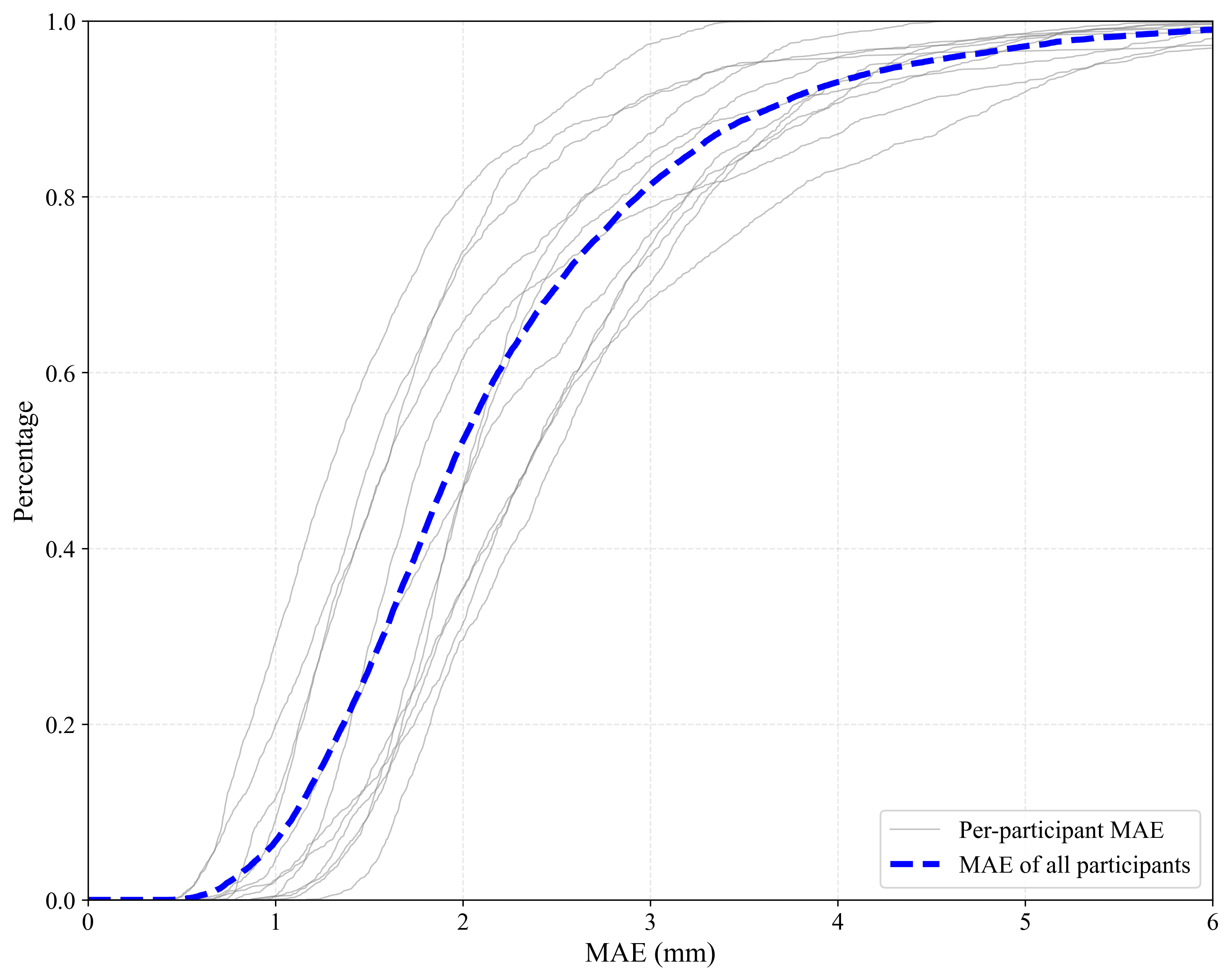}} \hfill
\subfloat[]{
		\includegraphics[width=0.27\textwidth, height=0.2\textheight]{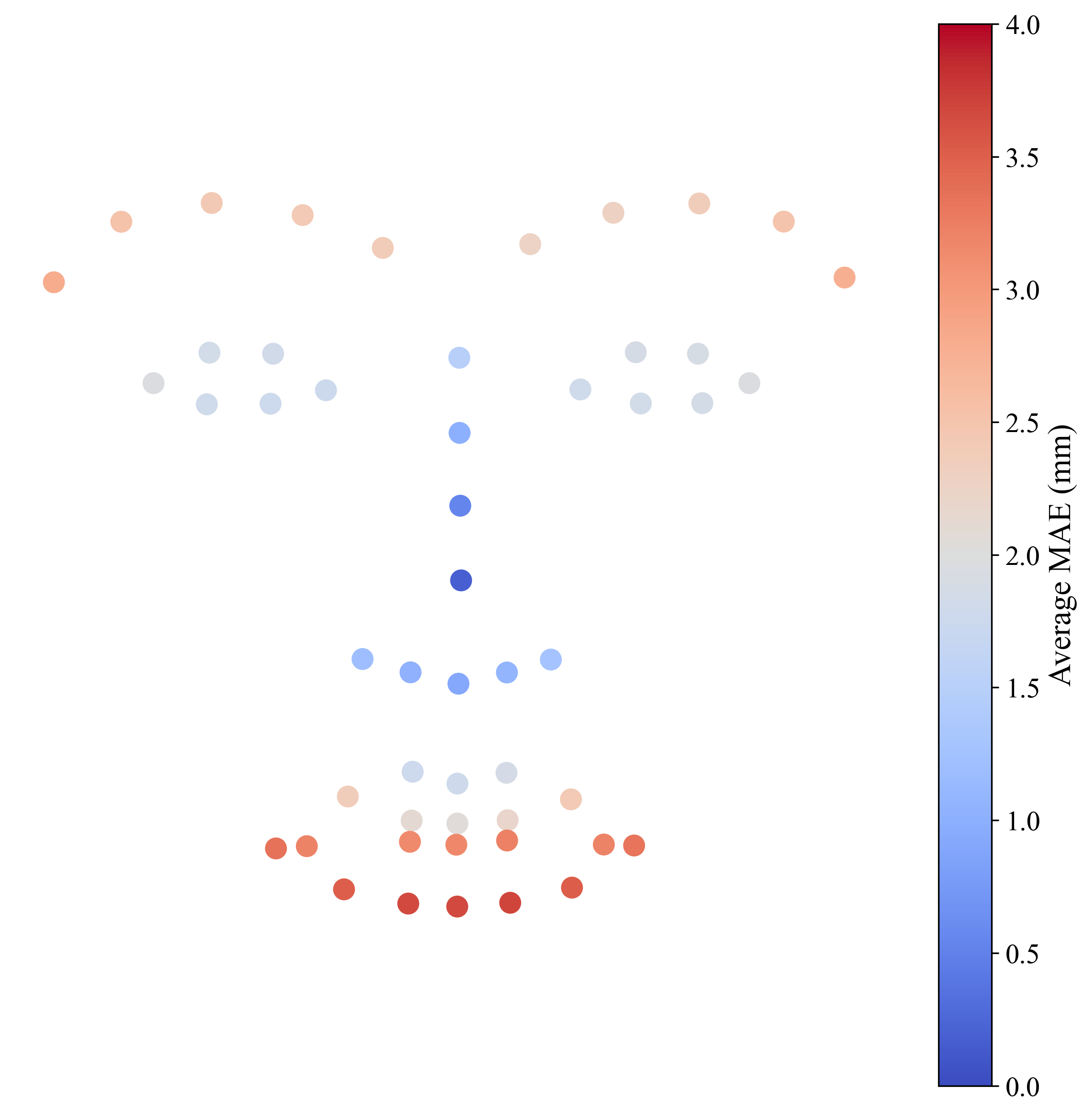}}
\caption{Performance of continuous facial landmark tracking for each participant. (a) Per-participant Landmark Tracking Error. (b) MAE CDF for each participant. (c) Visualization of average MAE.}
\label{fig:5}
\end{figure*}

\subsection{User-Dependent Evaluation}
For each user's 12 recorded sessions, we randomly selected five sessions for training, while the remaining sessions were used for testing in a randomized cross-validation manner. Fig.~\ref{fig:5}(a) presents the average Mean Absolute Error (MAE) and Normalized Mean Error (NME), along with their corresponding standard deviations, for all 51 facial landmarks of each participant. Specifically, the blue bars represent the average MAE (in millimeters), which reflects the absolute spatial deviation between the predicted and ground truth landmarks. The red bars represent the average NME (in percentage), which measures the relative deviation normalized by the facial size, enabling cross-user performance comparison regardless of individual facial dimensions. These two complementary metrics provide a comprehensive evaluation of both absolute and relative reconstruction accuracy. IMUFace achieves an average MAE of 2.21~mm and an NME of 3.40\%, with standard deviations of 0.99~mm and 0.90\%, respectively. Among all participants, U4 achieves the best performance with an MAE of only 1.46~mm, while U8 exhibits the highest MAE at 2.72~mm. Fig.~\ref{fig:5}(b) illustrates the Cumulative Density Function (CDF) of MAE for each individual participant as well as across all participants. The CDF curve reveals that 60.0\% of the samples have an MAE below 2.21~mm, indicating a high overall prediction accuracy. This distribution further confirms the robustness of our model in reconstructing facial landmarks with low error rates across different users. In addition, different landmarks tend to exhibit varying error levels due to their distinct movement patterns during facial expressions. Fig.~\ref{fig:5}(c) visualizes the average MAE for each of the 51 facial landmarks. We observe that landmarks located on the eyebrows and mouth tend to have higher reconstruction errors, while those around the eyes generally exhibit lower errors. This is mainly because the eyebrows and mouth undergo more dynamic deformations during expression changes, making their accurate reconstruction more challenging.

\subsection{User-Adaptive Evaluation}
\begin{figure}[ht]
    \centering
    \includegraphics[width=0.6\linewidth]{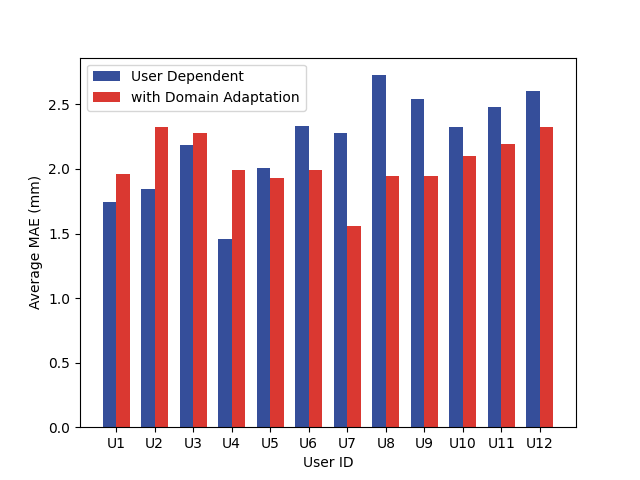}
    \caption{Domain adaptation.}
    \label{fig:6}
\end{figure}

Due to the need for improvement in cross-user testing results, we attempted to introduce domain adaptation techniques. First, we trained a model using data from other users, then fine-tuned the model's linear layers with 2-minute data from a specific user, and finally tested the model using the data from the same user. Fig.~\ref{fig:6} compares the performance differences between the user-dependent and domain-adapted models. It can be observed that the performance of the domain-adapted model is significantly improved compared to the user-dependent model, with an average MAE of 2.05 mm for the specific user. This improvement may be attributed to the unique characteristics of each user's facial muscle movements. Through fine-tuning, the model can learn specific patterns from the target user's data, thereby enhancing its prediction performance.

\begin{table}
    \caption{Comparison result of different Models.}
    \label{tab:modelselection}
    \centering
    \begin{tabular}{c c c}
        \hline
        \textbf{Models} & \textbf{Average MAE} & \textbf{Average NME}\\
        \hline
        Bi-LSTM~\cite{zhang2015bidirectional}           & 2.57 mm                 & 3.97\%   \\
        TCN~\cite{verma2021expressear}                  & 2.39 mm                 & 3.68\%   \\
        IF-ConvTransformer~\cite{zhang2022if}           & 2.27 mm                 & 3.50\%   \\
        Mamba2~\cite{dao2024transformers}               & 2.37 mm                 & 3.64\%   \\
        \textbf{IMUTwinTrans}                           & \textbf{2.21 mm}        & \textbf{3.40\%} \\
        \hline 
    \end{tabular}
\end{table}

\begin{figure}
    \centering
    \includegraphics[width=0.6\linewidth]{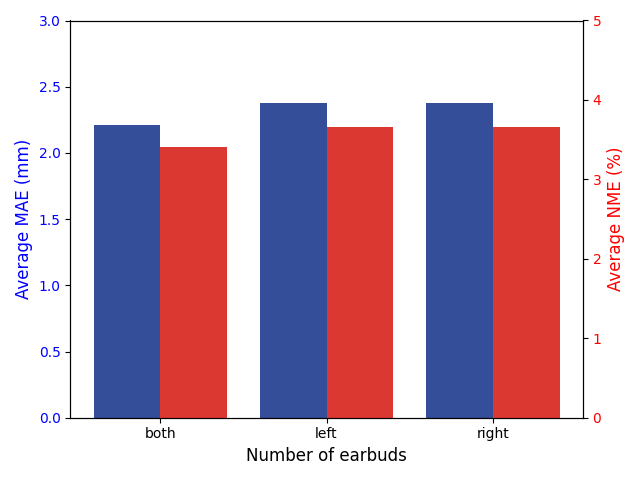}
    \caption{Number of earbuds.}
    \label{fig:66}
\end{figure}

\subsection{Micro-benchmark Tests}

\subsubsection{Model Selection}

To verify the effectiveness of our model, we compared it with several deep learning network methods commonly used for IMU data analysis to demonstrate the superiority of our model. We evaluated the performance of different networks under user-dependent experimental settings, and the results are presented in Table~\ref{tab:modelselection}.

\begin{figure}
    \centering
    \includegraphics[width=0.6\linewidth]{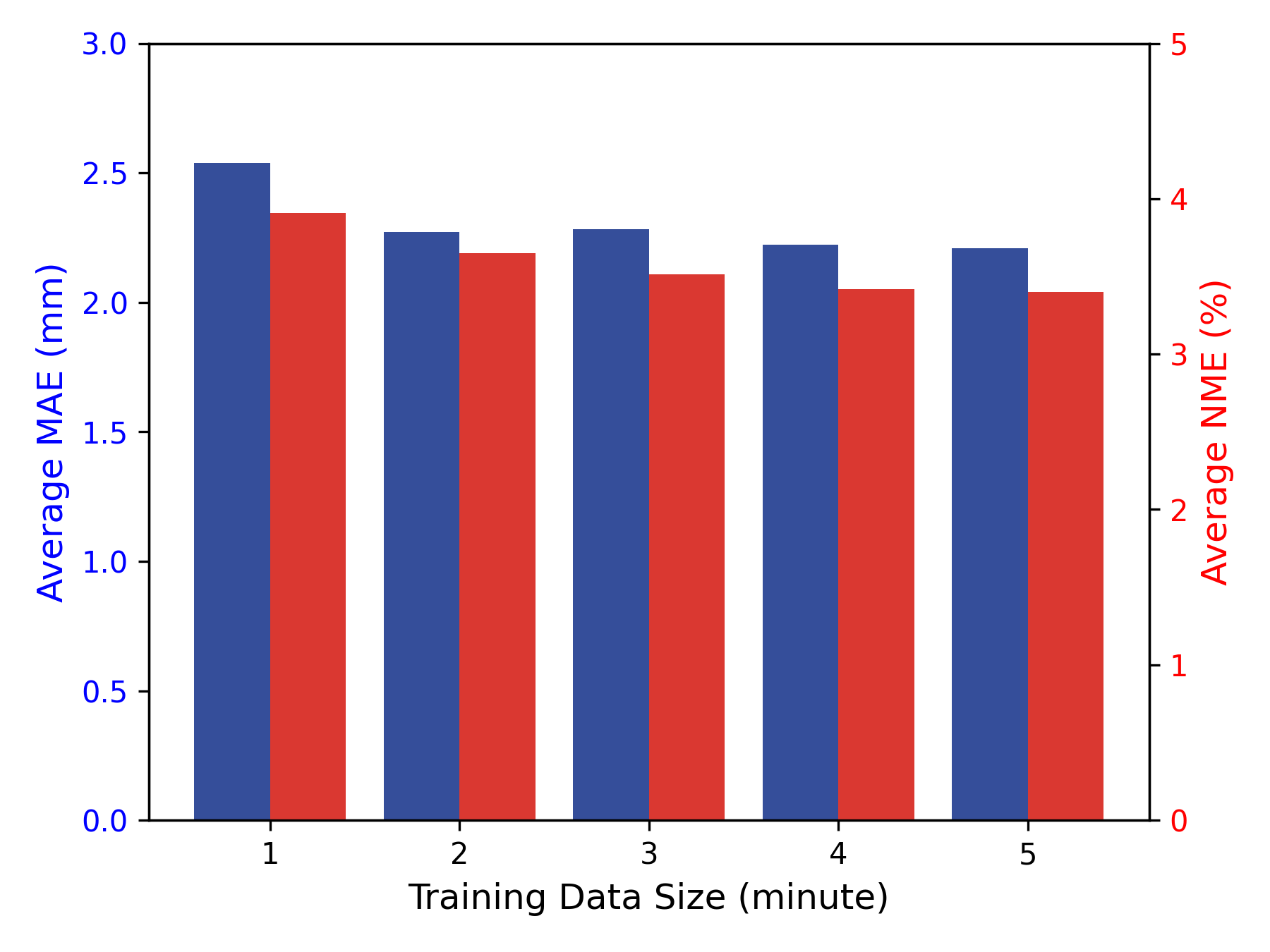}
    \caption{Training size vs error.}
    \label{fig:Training_duration_vs_error}
\end{figure}

\subsubsection{Number of Earphones}
To validate the system's efficacy under single-earphone conditions, we investigated the accuracy when using only a single earphone. As shown in Fig.~\ref{fig:66}, although using both earphones can provide the best accuracy by integrating sensor data from both ears, the accuracy with individual earphones (left or right) is also very close (2.40 mm, 2.36 mm).

\subsubsection{Impact of Training Data Size}
In practical applications, the amount of data that needs to be collected from users for training can affect user experience. Therefore, it is necessary to explore further the relationship between system performance and the required training data to minimize the burden of data collection on users. Fig.~\ref{fig:Training_duration_vs_error} shows the overall system performance when the training data size varies from 1 minute to 5 minutes for each participant. We observe that with 1 minute of training data, the accuracy already reaches 2.66 mm, and with 5 minutes of training data, the accuracy saturates at around 2.21 mm. It is shown that IMUFace can still achieve a reasonable level of accuracy even with a relatively short period of user data collection, which offers a high level of user-friendliness.

While the average MAE between two and five minutes of training data appears numerically close (2.27~mm vs. 2.21~mm), a more detailed analysis reveals a noticeable improvement in the normalized mean error (NME), which decreases from 3.65\% to 3.40\%. Since NME accounts for differences in face scale and user variation, this reduction suggests that the model trained with five minutes of data has better generalization across diverse users and face shapes. This improvement in normalized prediction accuracy justifies the use of a longer training duration. Additionally, the extended data collection enhances the model's robustness and stability in handling complex facial dynamics, making it more reliable for real-world applications.

\subsubsection{Comparison with Existing Solutions}
To compare the performance of IMUFace with existing vision-based and wearable methods, Table~\ref{tab:vision_comparison} presents the NME results across a range of representative systems. SDM~\cite{zhu2015face}, CFSS~\cite{xiong2013supervised}, and HRNet~\cite{wang2020deep} are classic vision-based approaches that rely on public image datasets such as 300-W to track 68 facial landmarks. In contrast, mm3DFace~\cite{xie2023mm3dface} and BioFace-3D~\cite{wu2021bioface} adopt alternative sensing modalities—millimeter-wave and EMG signals respectively—on self-collected datasets. EARFace~\cite{zhang2023earphone} is another non-visual method that uses in-ear sensing by emitting and receiving acoustic signals in the ear canal.

Our method, IMUFace, achieves an NME of 3.40\% on a self-collected dataset with 51 landmarks, which is very close to the performance of BioFace-3D (3.38\%). However, unlike BioFace-3D, which requires attaching electrodes directly to the user’s face, IMUFace relies only on lightweight wearable IMU sensors, making it significantly more comfortable and user-friendly.

Although IMUFace’s accuracy is slightly lower than EARFace (3.14\%), it avoids several practical limitations of the EARFace system. Specifically, EARFace relies on wired earphones and uses active acoustic signal emission into the ear canal, which may pose risks to user hearing health, involves higher power consumption, and compromises convenience due to its tethered design. In contrast, IMUFace offers comparable performance while being low-power, wireless, and non-invasive. These advantages highlight its strong potential for practical deployment, particularly in scenarios where visual input is unavailable or privacy-sensitive environments are involved.

\begin{table}[ht]
  \caption{Comparison with existing solutions.}
  \label{tab:vision_comparison}
  \centering
  \begin{tabular}{c c c c}
    \toprule
    \textbf{Methods} & \textbf{Dataset} & \textbf{Landmarks} & \textbf{NME (\%)} \\
    \midrule
    SDM~\cite{zhu2015face}           & 300-W           & 68  & 7.52 \\
    CFSS~\cite{xiong2013supervised}         & 300-W           & 68  & 5.76 \\
    HRNet~\cite{wang2020deep}       & 300-W           & 68  & 2.87 \\
    mm3DFace~\cite{xie2023mm3dface} & Self-collected  & 68  & 3.94 \\    
    BioFace-3D~\cite{wu2021bioface} & Self-collected & 53 & 3.38 \\
    EARFace~\cite{zhang2023earphone} & Self-collected   & 51 & 3.14 \\
    IMUFace (Ours)     & Self-collected & 51 & 3.40 \\
    \bottomrule
  \end{tabular}
\end{table}

\subsubsection{Latency and Power Consumption}
The inference time for the 51 facial landmarks was measured on a single NVIDIA GTX 4090 GPU. Our designed model requires only about 1.26 ms to reconstruct a single frame, which is sufficient for real-time applications.

In addition, we used a power monitoring device to measure the power consumption of IMUFace. Specifically, when the system is idle, that is, the microcontroller unit (MCU) is in sleep mode and does not transmit data via Bluetooth, the average power consumption of IMUFace is 50 mW (13.5 mA @ 3.70 V). When IMUFace actively collects IMU data and transmits it via Bluetooth, the total system power consumption is 58 mW (15.7 mA @ 3.70 V). This indicates that with a 100 mAh lithium polymer battery, IMUFace can continuously record data for up to 6.5 hours, meeting the needs of most applications. We examined the power consumption of recent studies on 3D facial reconstruction and summarized the overall power consumption of the system, including the use of sensing and communication power, in Table~\ref{tab:powerconsumption}. Our analysis shows that IMUFace consumes the least power among all sensing technologies evaluated, with energy consumption 58.0\% lower than that of the next most efficient technology, Bioface-3D~\cite{wu2021bioface}.

\begin{table}
  \caption{Comparison of System Power Consumption among Related Existing Studies}
  \label{tab:powerconsumption}
  \centering
    \begin{tabular}{c c c}
        \hline
        \textbf{Studies}&\textbf{Sensors}&\textbf{Power Consumption}\\
        \hline
        NeckFace~\cite{chen2021neckface}    & Cameras          & 4 W                \\
        C-Face~\cite{chen2020c}             & Cameras          & \textgreater 4 W    \\
        BioFace-3D~\cite{wu2021bioface}     & Biosensors       & 138 mW             \\
        EarIO~\cite{li2022eario}            & Acoustics        & 153.7 mW           \\
        EyeEcho~\cite{li2024eyeecho}        & Acoustics        & 167 mW             \\
        \textbf{IMUFace}                    & \textbf{IMUs}    & \textbf{58 mW}     \\
        \hline
    \end{tabular}
\end{table}

\section{Discussion}

This study explores the use of conventional motion sensors, such as accelerometers and gyroscopes, for capturing facial expression data, emphasizing their practicality and portability for a wide range of applications. A key challenge identified is the variability in facial movements across different users, which impacts the model's performance. To address this, incorporating a small amount of user-specific data has been shown to improve accuracy. This approach offers promising applications in fields like healthcare, gaming, and human-computer interaction by advancing real-time facial expression tracking and enabling innovative functionalities.

\subsection{User-independent Model}

Experimental results reveal that fine-tuning the model with user-specific data significantly enhances performance compared to a purely user-independent model. This improvement arises from the inherent variability in facial movement characteristics, such as intensity, responsiveness, and sensitivity, among individuals. These differences create unique data domains, posing challenges for generalized models. Although fine-tuning improves accuracy, it introduces a trade-off in usability, as new users must register and provide data for personalization.

To address this limitation, future work will focus on generating extensive virtual IMU datasets using multimodal inputs, such as video or depth sensors. By leveraging synthetic data, we aim to train a robust and generalized user-independent model capable of adapting to a broader user base with minimal personalization. Additionally, investigating techniques like domain adaptation and transfer learning could help bridge domain gaps, enhancing the model's generalization capabilities. These strategies aim to maintain high performance while improving the system's usability for diverse users.

\subsection{Effects of Body Movements}

While the system accounts for variations in initial head pose, dynamic body movements—such as repetitive head bounces during activities like walking or running—introduce significant noise into IMU signals~\cite{prakash2019stear}. This noise complicates the accurate detection of facial expressions, reducing the system's reliability in real-world scenarios. To mitigate this issue, future research will explore advanced filtering techniques designed to isolate and remove noise caused by body movements while preserving subtle facial expression signals. Methods such as adaptive filtering, wavelet transforms, or machine-learning-based noise suppression may improve signal quality. Furthermore, integrating complementary sensors, such as magnetometers or pressure sensors, could provide additional context to distinguish between facial and body-induced movements. These enhancements aim to ensure reliable system performance in dynamic and mobile environments.

\subsection{Practical Applications and Industry Implications}

The proposed methodology demonstrates significant potential across various sectors, highlighting its practical value. In healthcare, it enables remote monitoring of patients with neurological disorders, facilitating the assessment of facial muscle activity and therapy progress. In gaming and virtual reality, it offers camera-free, real-time facial expression tracking, enhancing user privacy and reducing hardware complexity. For human-computer interaction, it supports more intuitive communication between users and devices. In the automotive industry, the system can monitor driver fatigue or emotional states, improving safety and user experience. These diverse applications underline the method's versatility and its potential to address practical challenges across multiple disciplines.

\section{Conclusion}
This article proposes and designs IMUFace, a system that reconstructs the 3D facial expressions of users using IMUs that are attached to (and potentially built into) wireless earphones. The system can detect the motion changes associated with facial expressions and achieve user-friendly, low-effort, continuous 3D facial reconstruction. We extracted frequency-domain features from the data and designed a lightweight deep learning model, IMUTwinTrans, based on a ConvTransformer architecture to achieve 3D facial reconstruction by leveraging both temporal and frequency features of the IMU signal. It achieves an average facial landmark tracking accuracy of 2.21 mm across 12 users. The deep learning model inference time is only 1.26 ms, satisfying real-time requirements. Our device consumes only 58 mW of power during operation and can run for 6.25 hours on a 3.7 V, 100 mAh battery. Numerous applications in the areas of affective computing, emotional well-being monitoring, facial expression recognition, AR/VR, accessibility, and user interfaces can be further explored in future work. 

\bibliography{sample}
\end{document}